# Negative activation volume for dielectric relaxation in hydrated rocks

A.N. Papathanassiou *, I. Sakellis, and J. Grammatikakis

University of Athens, Physics Department, Section of Solid State Physics, Panepistimiopolis, 15784 Zografos, Athens, Greece

**Abstract**

Negative defect activation volumes are extremely rare in solids. Here, we report for the first time that this holds in a couple of hydrated rocks for dielectric relaxation by exploring the complex impedance spectra at various pressures and temperatures. The present findings mean that the relaxation time of the relevant relaxation mechanisms decreases upon increasing pressure, thus it may become too short at higher pressure and hence lead to the emission of transient electric signals before fracture. This may constitute the long-standing laboratory confirmation for the explanation of the generation of electric signals prior to an earthquake, as recently pointed out by Uyeda et al [Tectonophysics 470 (2009) 205-213].

The dielectric relaxation time $\tau$ is temperature and pressure dependent, since, according to the rate theory $\tau(P,T) = (\lambda \nu)^{-1} \exp(g^{act}/k_B T)$, where $\nu$ is the vibration frequency of the relaxing charge carrier, $\lambda$ is a geometrical constant, $g^{act}$ is the Gibbs activation energy for relaxation and $k_B$ is the Boltzmann's constant. Note that all thermodynamic quantities appearing in this paper are related to relaxation or, alternatively, to the localized motion of the electric charge entities that form the relaxing 'dipoles'. By differentiating with respect to pressure at constant temperature, we get:

$$(\partial \ln \tau / \partial P)_T = -\frac{\gamma}{B} + \frac{\upsilon^{act}}{k_B T} \qquad (1)$$

where $\gamma \equiv -(\partial \ln \nu / \partial \ln V)_T$ is the Grüneisen constant (V denotes the volume), $B \equiv -(\partial P / \partial \ln V)_T$ is the isothermal bulk modulus and $\upsilon^{act} \equiv (\partial g^{act}/\partial P)_T$ is the activation volume (Varotsos and Alexopoulos, (1986)). In ionic crystals doped with aliovalent impurities a number of extrinsic defects (e.g., vacancies, interstitials) are produced for reasons of charge compensation. A portion of these vacancies and/or interstitials may be located close to the aliovalent impurities, thus forming the aforementioned electric 'dipoles' (Varotsos and Miliotis, 1974) giving rise to dielectric relaxations. The remaining vacancies and/or interstitials are 'free' (in the sense that they lie far from the impurities), thus they may contribute to the dc conductivity (Varotsos and Alexopoulos, 1978; Varotsos and Alexopoulos (1979)). In particular, in alkaline earth fluoride crystals doped with monovalent (trivalent) rare earth ions, a number of dielectric relaxations were found and were ascribed to the rotation of impurity-vacancy (interstitial) dipole complexes (Varotsos and Alexopoulos (1980); Varotsos and Alexopoulos (1981)). Alkaline earth carbonate salts are representative materials of the earth's crust in Greece; calcite ($CaCO_3$) and magnesite ($MgCO_3$) are abundant in polycrystalline form (limestone and leukolite, respectively) and have been dielectrically characterized (Papathanassiou and Grammatikakis (1996); Papathanassiou and Grammatikakis, 1997; Papathanassiou, 1999 , 2001a).

Porosity in rocks contributes to the heterogeneity and gives rise to interfacial polarization. Intense polarization phenomena appear when porosity is filled (partially or fully saturated) with water (Varotsos, 2005; Papathanassiou, 2000a, b , 2001a, b; Endres and Knight, 1991; Nettelblad, 1996). In nature, the role of water in rocks (which is abundant in the earth's crust) is dual: (i) it enhances – as mentioned previously - the dielectric properties of the rock and (ii) induces the so-called *dilatancy* when the rock is pressurized. In particular, the latter occurs in the hypocenter area of an earthquake which is surrounded by water saturated porous rock with fluid-filled pore channels and the pre-earthquake stage is assumed to be accompanied by the appearance of fresh cracks in the fracture zone (Surkov et al, 2002). The time evolution of the stress field prior to an earthquake results in the dynamic compression of water-filled rocks (Morgan et al, 1989; Morgan and Nur, 1986).

Different polarization mechanisms occur in a rock: defect dipole rotation, interfacial polarization, double layer polarization etc. All these can be regarded as effective electric 'dipoles'. Electric 'dipoles' in rocks, which are influenced by polarizing fields such as the stress field which plays a role similar to that of an external field (Varotsos 2005), undergo a monotonic compression and their characteristic relaxation time is modified with increasing pressure. According to Eq. (1), the percentage variation of the relaxation time upon compression is determined by $\upsilon^{act}$. Negative activation volume values indicate that the rotational mobility of electric 'dipoles' is enhanced on increasing pressure. Negative activation volumes were experimentally found in a few materials, by studying the pressure dependence of the electrical conductivity (in NAFION hydrogels (Fontanella et al, 1996)) or the pressure dependence of the dielectric relaxation (in $\beta$-$PbF_2$ doped with lanthanum (Fontanella et al, 1982), and semi-conducting polypyrrole (Papathanassiou et al, 2006, 2007)). Varotsos, Alexopoulos and Nomicos (Varotsos et al, 1982, see also Varotsos and Alexopoulos 1986) in order to explain the generation of seismic electric signals (SES) (i.e., the electric signals that are observed (Varotsos, 2005; Varotsos at al, 2006a, b; Varotsos and Alexopoulos, 1984a, b; Varotsos et al 1996, Varotsos et al, 2003a, b; Varotsos et al, 2002) before earthquakes), suggested the following theoretical model, which according to Uyeda et al (2009) is unique, but yet lacks laboratory confirmation; they treated the earth's crust as a solid rich in 'dipoles' in a polarizing field (i.e., the mechanical stress field). *They assumed that negative activation volumes probably exist in rocks* and, based on this assumption, asserted that the monotonic increase of pressure during the earthquake preparation, reduces the relaxation time of these 'dipoles'. Hence, at a certain critical pressure (smaller than that at which rupture occurs) the relaxation time of 'dipoles' becomes too short and then the 'dipoles' undergo a transition from random-orientation state to an oriented one. The time variation of the polarization of the rock yields the emission of a transient polarization current. The latter idea is known in solid state physics as pressure stimulated currents (PSC) (for a brief review of PSC see also Varotsos et al 1998). Thus, in short, PSC can simulate the appearance of SES *under the assumption that negative activation volumes for 'dipole' relaxation exist* in hydrated rocks. The present paper constitutes a laboratory investigation of the validity of the latter assumptions.

Here, we studied high purity leukolite ($MgCO_3$) from Euboea, Greece (Papathanassiou, 1999) and kataclastic limestone (polycrystalline $CaCO_3$) from Greece (Papathanassiou, 2001a) that are commonly found in faults, where earthquakes are most likely to occur. A Novocontrol (Germany) high pressure dielectric spectroscopy system (0.3 GPa is the maximum hydrostatic pressure), was used where silicon oil is the pressure transmitting fluid. Metallic electrodes were pasted on to the parallel surfaces of the specimen and a very thin insulating layer of epoxy covered the specimen to prevent contamination from the pressure transmitting fluid (Papathanassiou and Grammatikakis, 1996; Papathanassiou, 1998, 2001, 2002). Rock samples were hydrated immersing them in distilled water at $70^0$ C for 4 days. The elevated temperature yield expansion of the specimen and opening of the pore cavities and, at the same time, diffusivity of water molecules is enhanced. It was observed that 24 h was an adequate time interval to achieve saturation. Following this procedure, leukolite and limestone absorbed 1.0 and 0.6 wt % water, respectively. The tangent loss angle function $\tan\delta \equiv Im(\varepsilon)/Re(\varepsilon)$, where $Im(\varepsilon)$ and $Re(\varepsilon)$ are the imaginary and the real part of the complex permittivity $\varepsilon$ (relative to its value of free space) traces a dielectric relaxation mechanism in the frequency domain by exhibiting a maximum at a frequency $f_{max,\tan\delta} = \sqrt{\varepsilon_s/\varepsilon_\infty}\tau^{-1}$, where $\varepsilon_s$ and $\varepsilon_\infty$ denote the static and high-frequency (relative) permittivity, respectively. These values can be obtained by extrapolation of the available broadband complex impedance data to the low and high frequency limit, respectively. By differentiating the natural logarithm of the later equation with respect to pressure at constant temperature, we get:

$$\left(\partial \ln\tau/\partial P\right)_T = \tfrac{1}{2}\left(\partial \ln(\varepsilon_\infty/\varepsilon_s)/\partial P)\right)_T - \left(\partial \ln f_{max,\tan\delta}/\partial P\right)_T \qquad (2)$$

From Eqs (1) and (2), we get the activation volume by studying the pressure variation of $f_{max,\tan\delta}$ and the pressure variation of $\varepsilon_\infty/\varepsilon_s$:

$$\upsilon^{act} = k_B T\left\{\frac{\gamma}{B} + \frac{1}{2}(\partial \ln(\varepsilon_\infty/\varepsilon_s)/\partial P))_T - (\partial \ln f_{max,\tan\delta}/\partial P)_T\right\} \qquad (3)$$

Isotherms of tanδ vs frequency of hydrated leukolite for different pressure values are depicted in Fig. 1. In the same diagram, a typical relaxation spectrum for as-received (not hydrated) leukolite specimen is presented, indicating that the relaxation mechanism in hydrated leukolite is related with the presence of water in the porosity. The temperature dependence of the peak maxima is depicted in the inset of Fig. 1, where $\ln f_{max,tan\delta}$ is plotted against $1/k_B T$ for P=155MPa. By using Eq. (3) we get a negative activation volume (Table 1). The error shows that $\upsilon^{act}$ is undoubtedly negative, e.g., the relaxation time decreases on pressurization according to Eq. (1). In Fig. 2 we see isotherms of tanδ vs frequency of hydrated limestone for different pressure values. A broad relaxation appearing in the low frequency region has positive activation volume, while, another one appearing in the high frequency region of the dielectric spectra has negative activation volume (Table 1). It is worth noticing that the (absolute) value of the activation volume of leucolite is rather large and the activation enthalpy (obtained from the slope of the line presented in the inset of Fig.1) is $(0.90 \pm 0.05)$eV and the pre-exponential factor of the Arrhenius relation is of the order of $10^{-13}$ s. It is difficult to correlate the observed relaxation mechanisms with specific microscopic processes due to the complexity of rocks. Different scenarios can explain the appearance of the relaxation peaks, such as defect dipole rotation, interfacial polarization, dynamics of water molecules in confined areas or surfaces, etc. The potential conclusion drawn from Table 1, which is important, as mentioned above, for the explanation of the SES generation, is that negative $\upsilon^{act}$ values for 'dipole' relaxation are obtained in hydrated alkaline earth carbonate rocks, regardless the type of the relaxing 'dipoles' (Varotsos, 2005; pp. 230-231 and p. 258).

In summary, the relaxation mechanisms related with the presence of the saturating water in leukolite and limestone exhibit negative activation volumes for dielectric relaxation. Thus, these hydrated rocks provide characteristic examples of the rare cases where negative volumes for defect activations have been observed to date. Furthermore, the present finding is a novel validation of the theoretical model, which has been suggested as underlying physics for the generation of transient electric signals long before an earthquake occurrence.

**Table 1** Estimation of the activation volumes evaluated from complex impedance data in the frequency domain, by using Eq. (3). The ratio γ/B was estimated by considering

a value γ=1.7, which is a typical value for ionic crystals, and the bulk modulus B was taken from (Wang, (1974). LF and HF label the low frequency and high frequency mechanisms of limestone, respectively.

|  |  | T(K) | $\gamma/B$ (GPa$^{-1}$) | $(\partial \ln(\varepsilon_\infty/\varepsilon_s)/\partial P)$ (GPa$^{-1}$) | $(\partial \ln f_{max, \tan \delta}/\partial P$ (GPa$^{-1}$) | $\upsilon^{act}$ (Å$^3$) |
|---|---|---|---|---|---|---|
| Hydrated Leukolite |  | 349 | 0.0221 | -4.0 ± 0.5 | 21 ± 4 | -100 ± 25 |
| Hydrated Limestone | *LF* | 312 | 0.0884 | -7.6 ± 0.7 | -16 ± 1 | +5.0 ± 0.6 |
|  | *HF* |  |  |  | 4.2 ± 0.4 | -1.3 ± 0.2 |

References

Endres A.L. and Knight R.J. (1991), The effects of pore scale fluid distribution on the physical properties of partially saturated tight sandstone, Appl. Phys. 69, 1091-1098

Fontanella J. J., Wintersgill M.C., Figueroa D.R., Chadwick A.V. and Andeen C.C. (1982), Anomalous pressure dependence of dipolar relaxation inj rare earth doped lead fluoride, Phys. Rev. Lett. 51, 1892-1895

Fontanella J. J., Edmondson C. A., Wintersgill M. C., Wu Y. and Greenbaum S. G. (1996), High-Pressure Electrical Conductivity and NMR Studies in Variable Equivalent Weight NAFION Membranes, Macromolecules 29, 4944-4951 doi: 10.1021/ma9600926

Morgan F. D., Williams E.R. and Madden T.R. (1989), Streaming potential properties of westerly granite with applications, J. Geophys. Res. 94, 12449-12461

Morgan F.D. and Nur A. (1986), EOS Trans. AGU 67, 1203

Nettelblad B. (1996), Electrical and dielectric properties of systems of porous solids and salt-containing nonpolar liquids, J. Appl. Phys. 79,7106-7113 doi:10.1063/1.361480

Papathanassiou A.N. and Grammatikakis J. (1996), Defect dipole relaxation in polycrystalline dolomite ($CaMg(CO_3)_2$), Phys.Rev. B 53, 16252-16257 doi: 10.1103/PhysRevB.53.16252

Papathanassiou A.N. and Grammatikakis J. (1997), The dielectric relaxation of the calcite type carbonate salts: Defect structure and defect dipole dynamics in polycrystalline magnesite, Phys. Rev. B 56, 8590-8598 doi: 10.1103/PhysRevB.56.8590

Papathanassiou A.N. (1999), Thermal depolarisation studies in leukolite (polycrystalline magnesite, $MgCO_3$),J. Phys. Chem.Solids 60, 407-414 doi: 10.1016/S0022-3697(98)00270-4

Papathanassiou A.N. and Grammatikakis J., (2000), Dielectric characterization of the water-matrix interaction in porous materials by Thermal Depolarization spectroscopy, Phys.Rev. B 61, 16514-16521 doi: 10.1103/PhysRevB.61.16514

Papathanassiou A.N. (2000b), Investigation of the dielectric relaxation and the transport properties of porous silicates containing humidity, Phys.:Condensed Matter 12, 5789-5800 doi: 10.1088/0953-8984/12/26/324

Papathanassiou A.N. (2002a), Evaluation of the relaxation parameters of interfacial polarization processes in calcite containing dolomite and quartz inclusions by TSDC spectroscopy, J. Phys. D: Appl. Phys. 34, 2825-2829 doi: 10.1088/0022-3727/34/18/317

Papathanassiou A.N. (2001b), On the polarization mechanisms related to the liquid-solid interaction, J. Phys. Condens. Matter 13, L791-L782 doi: 10.1088/0953-8984/12/26/324

Papathanassiou A.N. and Grammatikakis J. (1996b), Pressure variation of the electrical conductivity of dolomite [$CaMg(CO_3)_2$, Phys.Rev.B 53,16247-16251 doi: 10.1103/PhysRevB.53.16247

Papathanassiou A.N. (1998), Effect of hydrostatic pressure on the electrical conductance of polycrystalline magnesite ($MgCO_3$), Phys. Rev. B 58, 4432-4437

Papathanassiou A.N. (2001c), Pressure variation of the conductivity in single crystal calcite, Phys. Stat. Solidi (b) 228, R6-R7

Papathanassiou A.N. (2002b), Dependence of the electrical conductivity and the low-frequency dielectric constant upon pressure in porous media containing a small quantity of humidity, Electrochim. Acta 48, 235-239 doi: 10.1016/S0013-4686(02)00619-9

Papathanassiou A.N., Sakellis I. and Grammatikakis J. (2006), Separation of electric charge flow mechanisms in conducting polymer networks under hydrostatic pressure, Appl. Phys. Lett. 89, 222905 doi: 10.1063/1.2768623

Papathanassiou A.N., Sakellis I. and Grammatikakis J. (2007), Migration volume for polaron dielectric relaxation in disordered materials, Appl. Phys. Lett. 91, 202103 doi: 10.1063/1.2812538

Surkov V.V., Uyeda S., Tanaka H. and Hayakawa M. (2002), Fractal properties of medium and seismoelectric phenomena, J. Geodynamics 33, 477-487

Uyeda S., Nagao T. and Kamogaua M. (2009), Short-term earthquake prediction : Current status of seismo-electromagnetics Tectonophysics 470, 205-213

Varotsos P. and Miliotis D. (1974), New aspects on dielectric properties of alkali halides with divalent impurities, J.Phys.Chem.Solids 35, 927-930

Varotsos P. and Alexopoulos K. (1978), Curvature in the conductivity plots of silver halides as a consequence of anharmonicity, J. Phys. Chem. Solids 39, 759-761

Varotsos P. and Alexopoulos K.(1979), Possibility of the enthalpy of a Schottky defect decreasing with increasing temperature, J. Phys. C: Solid State 12, L761-L764.

Varotsos P. and Alexopoulos K. (1980), Migration entropy for the bound fluorine motion in alkaline earth fluorides, J. Phys. Chem. Solids 41, 443-446

Varotsos P. and Alexopoulos K. (1981), Migration parameters for the bound fluorine motion in alkaline earth fluorides, J. Phys. Chem. Solids 42, 409-410

Varotsos P.A. and Alexopoulos K.D. (1986), Thermodynamics of Point Defects and Their Relation with Bulk Properties, Editors: S. Amelinckx, R. Gevers and J. Nihoul, p. 79 and pp. 130-131, North-Holland, Amsterdam

Varotsos P., Eftaxias K., Lazaridou M., Antonopoulos G., Makris J. and Poliyiannakis J. (1996), Summary of the five principles suggested by P. Varotsos et al (1996) and the additional questions raised in his debate, Geophys. Res. Lett. 23, 1449-1452

Varotsos P., Alexopoulos K. and Nomicos K. (1982), Comments on the physical variation of the Gibbs energy for bound and unbound defects, Phys. Stat. Sol. (b) 111, 581-590

Varotsos P. and Alexopoulos K. (1984a), Physical properties of the variation of the electric field of the earth preceding earthquakes, Tectonophysics 110, 73-98

Varotsos P. and Alexopoulos K. (1984b), Physical properties of the variation of the electric field of the earth preceding earthquakes: Determination of the epicenter and magnitude, Tectonophysics 110, 99-125

Varotsos P., Sarlis N., Lazaridou M. and Kapiris P. (1998), Transmission of stress induced electric signals in dielectric media, J. Appl. Phys. 83, 60-70.

Varotsos, P.A., Sarlis, N.V. and Skordas, E.S. (2002), Long range correlations in the signals that precede rupture, Physical Review E 66, 011902, DOI: 10.1103/PhysRevE.66.011902

Varotsos, P.A., Sarlis, N.V. and Skordas, E.S. (2003a), Long-range correlations in the electric signals that precede rupture: Further investigations, Phys. Rev. E 67,021109, DOI:10.1103/PhysRevE.67.021109.

Varotsos, P.A., Sarlis, N.V. and Skordas, E.S. (2003b), Attempt to distinguish electric signals of a dichotomous nature, Phys. Rev. E. 68, 031106, DOI: 10.1103/PhysRevE.68.031106.

Varotsos P.A. (2005), The Physics of the Seismic Electric Signals, TERRAPUB, Tokyo

Varotsos P.A, Sarlis N.V., SkordasE.S. and Lazaridou M. S. (2005a), Natural entropy fluctuations discriminate similar looking signals emitted from systems of different dynamics, Phys. Rev. E 71, 011110 doi 10.1103/PhysRevE.71.011110

Varotsos P.A., Sarlis N.V., Tanaka H.K., and Skordas E.S. (2005b), Some properties of the entropy in the natural time, Phys. Rev. E 71, 032102 doi: 10.1103/PhysRevE.71.032102

Varotsos P. A., Sarlis N. V., Skordas E., Tanaka H.K. and Lazaridou M.S. (2006a), Attempt to distinguish long-range temporal correlations from the statistics of the increments by natural time analysis Phys. Rev. E 74, 021123 doi 10.1103/PhysRevE.74.021123

Varotsos P. A., Sarlis N. V., Skordas E., Tanaka H.K. and Lazaridou M.S. (2006b), Entropy of seismic electric signals: Analysis in natural time under time reversal, Phys. Rev. E 73, 031114 doi 10.1103/PhysRevE.73.031114

Wang Chi-Yuen (1974), Pressure coefficient of compressional wave velocity for a bronzite, J. Geophys. Res. 79, 771-772

**Figure Captions**

**Figure 1.** Isotherms (349K) of tanδ vs frequency of hydrated leukolite. The dash line is a typical measurement on an as-received (not hydrated) specimen at 178 MPa. In the inset diagram, $\ln f_{max,tan\delta}$ is plotted against $1/k_B T$ for P=155MPa. The straight line fits the data points.

**Figure 2.** Isotherms (312K) of tanδ vs frequency of hydrated kataclastic limestone. The straight lines were drawn to indicate the variation of a couple of maxima upon pressure. From top line to bottom: Ambient pressure, 26 MPa, 53 MPa, 67 MPa, 85 MPa, 110 MPa, 140 MPa, 200 MPa, 250 MPa and 300 MPa , respectively.

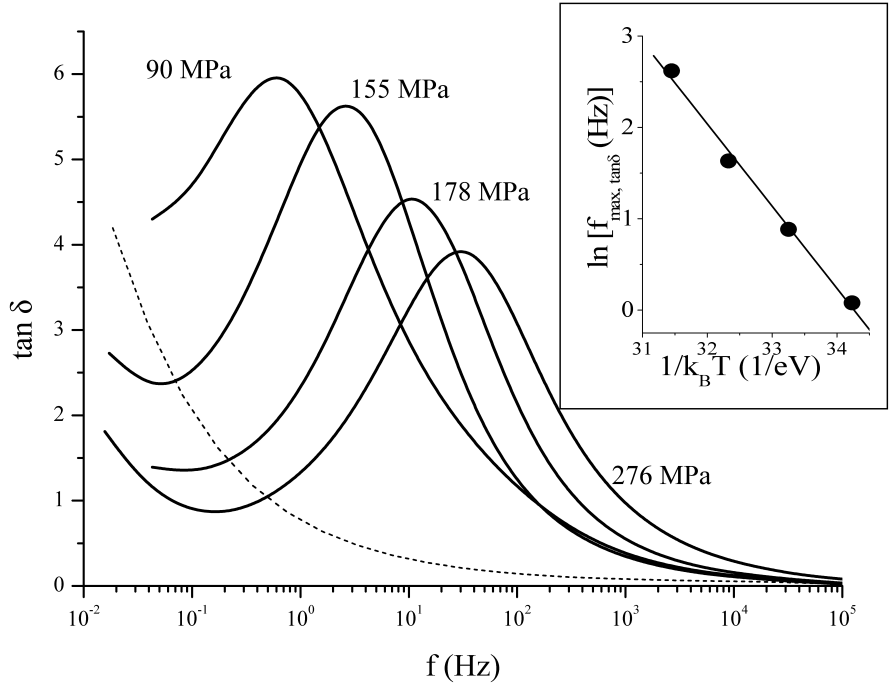

FIGURE 1

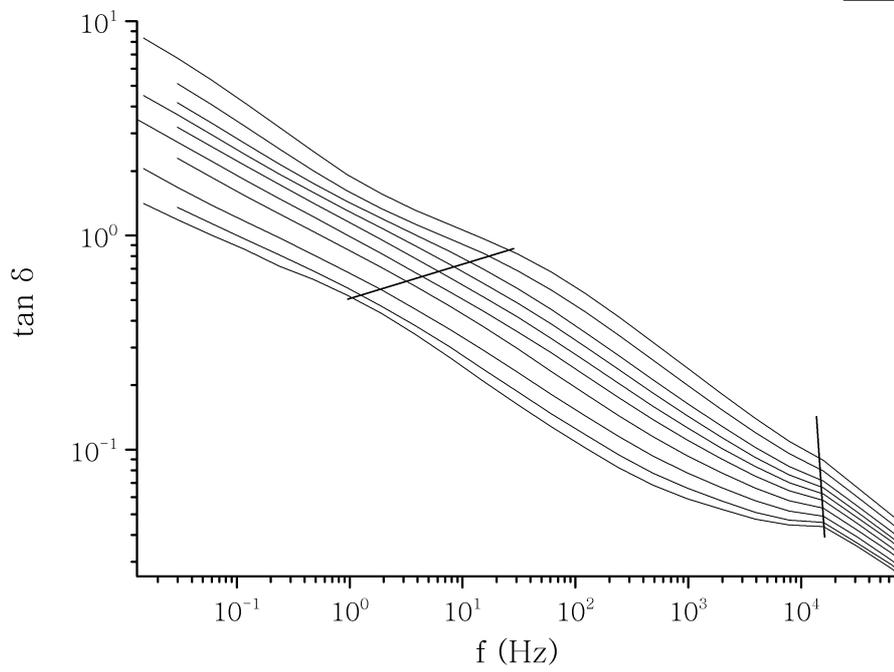

FIGURE 2